# License Compliance in Open Source Cybersecurity Projects

Ahmed Shah, Selman Selman, and Ibrahim Abualhaol

“ *One bad apple can spoil the bunch.* ”

Proverb

Developers of cybersecurity software often include and rely upon open source software packages in their commercial software products. Before open source code is absorbed into a proprietary product, developers must check the package license to see if the project is permissively licensed, thereby allowing for commercial-friendly inheritance and redistribution. However, there is a risk that the open source package license could be inaccurate due to being silently contaminated with restrictively licensed open source code that may prohibit the sale or confidentiality of commercial derivative work. Contamination of commercial products could lead to expensive remediation costs, damage to the company's reputation, and costly legal fees. In this article, we report on our preliminary analysis of more than 200 open source cybersecurity projects to identify the most frequently used license types and languages and to look for evidence of permissively licensed open source projects that are likely contaminated by restrictive licensed material (i.e., containing commercial-unfriendly code). Our analysis identified restrictive license contamination cases occurring in permissively licensed open source projects. Furthermore, we found a high proportion of code that lacked copyright attribution. We expect that the results of this study will: i) provide managers and developers with an understanding of how contamination can occur, ii) provide open source communities with an understanding on how they can better protect their intellectual property by including licenses and copyright information in their code, and ii) provide entrepreneurs with an understanding of the open source cybersecurity domain in terms of licensing and contamination and how they affect decisions about cybersecurity software architectures.

## Introduction

There are many types of open source cybersecurity packages that developers can leverage for product development and include within their proprietary products. Examples include penetration testing software tools that assist with identifying vulnerabilities and intrusion detection tools that are used to detect cyber-attacks. However, whether or not an open source package can be included within a commercial product will depend on the package license and the extent to which it restricts commercial activities such as the sale of the software and keeping derivative code confidential.

For the purposes of this article, we divide licenses into two categories: permissive and restrictive. The permissive category includes commercial friendly licenses, such as BSD, Apache, and MIT. In contrast, the restrictive category includes comparatively commercial unfriendly licenses, such as the GPL, that restrict the sale of software that includes an open source package with such a license.

Intellectual property and legal compliance issues can arise when companies fail to implement a thorough license evaluation process when they consume open source. The challenge is accentuated by the absence of a forced to click “I agree” to the license terms before installing or using code (Gaff and Ploussios, 2012). Contamination could occur when restrictively licensed code is copied into a permissively licensed project package or when a restrictively licensed package is copied into a permissively licensed project. Developers that are working under tight deadlines can easily over-look the licensing commitments of what they consume unless they have policies and tools in place to prevent contamination (Khanafer, 2015). This ease of consumption increases the risk of contamination. Developers need to know whether they are consuming code that is permissively licensed (i.e., commercial friendly) or restrictively licensed (i.e., commercial unfriendly). In the simplest case, they can simply check by inspecting the package license (i.e., the "readme" file or the "LICENSE" file), but complications can arise if a permissively licensed project silently hides code licensed under an incompatible restrictive license. However, to reduce the risk of contamination, developers should not assume that the same degree of diligence has been undertaken by other developers who contributed code to another project that is now being consumed as part of a separate package. Projects must take care not to inherit the problems of others, which can spread across projects in a viral manner as the code is copied. In a 2007 survey by Saugatuck Technology, 21% of respondents felt security/open/community concerns could inhibit the adoption of open source, while 12% felt that licensing issues and risks were a concern (Cited in Hassin, 2007). Failure to adhere to the open source licensing terms can lead to costly litigation, damage to a company's reputation, and cost spent to remediate contaminated code. For example, in 2009, the Software Freedom Conservancy, Inc. brought legal action for copyright infringement against 14 commercial electronics distributors including Westinghouse Digital Electronics, Best Buy, and Samsung (Klasfelld, 2011). These companies distributed code from BusyBox (an open source tool) in their products without adhering to the BusyBox license. The license stated that inheritors of BusyBox must make their own source code available to the public. Thus, licensing and copyright violations, in many cases resulting from code contamination, are substantial issues affecting vendors of software that leverages open source projects.

Within the cybersecurity domain, we investigated the extent to which projects with permissive open source package licenses (i.e., LICENSE and README files that refer to Apache, BSD, or MIT) are contaminated with a restrictive licensed file (GPL v1.0, 2.0, 3.0 ext). We examined more than 200 open source cybersecurity projects as an initial, exploratory study. By studying code contamination in open source cybersecurity projects and providing related insights about how contamination can be avoided, we ultimately seek to help developers make clean and profitable products.

Our motivation for analyzing the cybersecurity open source domain over other open source software domains comes from the authors' cybersecurity research exploring what tools can be used to create cybersecurity products and what cybersecurity tools can support or differentiate non-cybersecurity software product offerings. In addition, cybersecurity tools are vastly varied in type and function. Such tools include cyber-threat intelligence-sharing tools, software-defined radio tools, vulnerability and exploitation tools, and anti-virus tools.

This article is divided into four sections. First, we review the literature on open source licensing; how open source licensing can influence architecture; and how restrictive license contamination can lead to litigation. Next, we outline the origins of the sample projects that were studied and the analytical methods used. Then, we present our results, including information on the license types found in the study; the coding languages used; how well intellectual property rights are claimed in the sample; cases where restrictive licenses contamination occurred in permissively licensed packages; and proposed areas for future work. We conclude with a summary of results and recommendations.

## Literature Review

Restrictive licenses are generally considered "viral" because they require a consumer of the licensed code to distribute their own derivative source code under that same license. "Proprietary code distributed with or alongside GPL-licensed [open source software] as part of a larger program or application can in many cases be deemed a "covered work" along with the [open source software]. This means that the entire covered work – the proprietary code and OSS – can only be distributed under the GPL license terms" (Gaff and Ploussios, 2012). Permissive licenses allow consumers of the open source project to redistribute or sell the compiled binary without the need to expose any code to the public. Generally speaking, restrictive licenses allow consumers of the open source project to redistribute the compiled binary under the condition that the source code of the binary must be made available to the public and that the binary and source code cannot be sold. Not adhering to license terms in open source software could results in a copyright infringement claim or breach of contract, which may in turn lead to prohibition of further sales, impoundment and destruction of combined software, and legal fees (Gaff and Ploussios, 2012).

# License Compliance in Open Source Cybersecurity Projects

*Ahmed Shah, Selman Selman, and Ibrahim Abualhaol*

Most licenses are reciprocal licenses meaning they force all derived works to be licensed under the same license associated with the original copy of the component (Link, 2011). The General Public License (GPL; gnu.org/licenses/gpl.html) is the most common and notable example. Permissive licenses such as MIT (opensource.org/licenses/MIT) and Apache (apache.org/licenses/) have fewer restrictions and generally do not require the user to distribute their own derived work. Due to the variation of terms in each license type, licenses can be incompatible with each other if they are within the same open source package. In other words, if a developer is considering using multiple types of licensed open source projects, there is a risk that the licenses will not be compatible and that software therefore cannot be combined (Lokhman et al., 2013). For instance, a package that is licensed with Apache 2.0 is not compatible with GPL 1.0. Therefore, GPL 1.0 code should not exist in the Apache-licensed package's code base. In this article, the terms “license conflict” or “contamination” refer to a project with a permissive license contains restrictively licensed code.

*Open source licensing influencing architecture*

We define derived work as the result of enhancing or editing open source software. Depending on developer intentions, either to distribute derived work or publish their work while maintaining copyright ownership, open source legality and licensing issues must be faced. One approach, used by the Linux kernel, is the "core-periphery pattern" (Lokhman et al., 2013). The core of the Linux kernel owns the copyright for the core system, while applications built around this system (i.e., on the periphery) can be replaced with different applications to allow any number of versions, or distributions, of Linux to be created for different purposes and systems. This approach allows for license-compatible customization, and thereby enables usability, scalability, and modularity.

The main problem facing commercial companies are the obligations associated with the derived work (Hammouda et al., 2010). First, they must be aware of the licenses of the different components used in their systems, and second, they must make sure all these licenses are compatible. However, in some cases, it is hard to find a suitable project that has the appropriately compatible licenses and, therefore, software architecture considerations arise.

Conflicts can prohibit the integration of open source components and require extra effort to understand the limitations of the licenses used (Link, 2011). Consider the difference between the Lesser General Public License (LGPL; gnu.org/copyleft/lesser.html) and the GPL license (gnu.org/licenses/gpl.html). For code under the LGPL, the user is permitted to link it dynamically to other components without violating or enforcing the LGPL (Lokhman et al., 2013). In contrast, this same scenario with the GPL requires a separate executable if the software code is not being released. Thus, this requirement of the GPL can affect the architecture of the entire system, particularly when there is a mix of proprietary and open source components. For example, instead of linking components with the GPL component through control-driven communication, data-driven relationships must be used instead (Hammouda et al., 2010). Another approach is to use the "isolation pattern", which separates components from each other to avoid license conflicts (Hammouda et al., 2010). Depending on the nature of the system (i.e., hosted, distributed, released as open source), the system architecture must appropriately accommodate licensing obligations.

*Contamination leading to litigation*

There are many ways that a company's product can end up containing restrictively licensed source code, potentially triggering GPL-related litigation. Common violations include not distributing the source code of derivative works or failing to add appropriate copyright information or licenses to derivative works.

Many GPL contamination cases that lead to litigation often go through the following process:

1. *Release:* a third-party developer creates original source that is released under GPL.

2. *Contamination:* a commercial entity "consumes" the GPL code and (knowingly or unknowingly) adds the code to their commercial product.

3. V*iolation:* the commercial entity releases their GPL-contaminated product while not adhering to GPL terms (i.e., they fail to make their own source code available to the public).

4. *Indictment:* a company takes legal action against the GPL violator for not complying with GPL terms.

5. *Resolution:* the outcome of litigation.

The outcomes of litigation can be substantial, including but not limited to:

- Reputational damage

## License Compliance in Open Source Cybersecurity Projects

*Ahmed Shah, Selman Selman, and Ibrahim Abualhaol*

- Exposing customers to liability
- Threats of patent infringement for code tied to patents
- Making proprietary code open source
- Statutory damages
- Remediation costs of re-writing code

## Research Method

The source code for over 200 cybersecurity projects were downloaded which included tools for penetration testing, forensic investigation, intrusion detection, visualization, and network monitoring. We developed our dataset of projects by sampling a subset of the tools listed from the following three primary sources:

1. Kali Linux OS distribution (Offensive Security, 2015)

2. Department of Homeland Security's list of open source cybersecurity software (DHS, 2012)

3. Security Onion Linux OS distribution (Burks, 2015)

An overall dataset of 334 open source cybersecurity projects was created and three levels of analysis was conducted:

1. *Attribution:* Across all 334 projects, we determined the extent of: i) copyright information in each file, ii) license information in each file, iii) no copyright or license attribution in each file. The purpose of this analysis was to determine how many files across all projects have no copyright or license attribution and also what types of license attribution were applied to a file, if any.

2. *License conflicts:* Out of the 334 open source cybersecurity tools that we downloaded, tools for which we could confirm the package license from the project's website or from the source code's package license (i.e., "COPYING" or "LICENSE" file) were selected for an analysis of license conflicts. The resulting subset of 255 projects were examined for evidence of GPL file contamination in a permissively licensed package. To look for patterns in the appearance of license conflicts, we evaluated license conflicts against the number of lines of code per package and the types and number of coding languages used in each of these projects.

3. *Third-party code:* Across a sample of 243 projects where we could confirm the licenses, we assessed the volume of third-party code as a proportion of the total code volume in each project. To look for patterns in the appearance of license conflicts, we evaluated license conflicts against the lines of code per package.

To conduct the three levels of analysis described above, we scanned and analyzed the downloaded software packages using Protecode's Enterprise System 4 code-scanning engine (protecode.com/our-products/system-4/). The analyses included determination of the number of lines of code per package; likely third-party volume per package; license type per package; programming languages used per package; if a copyright or license existed in a code file; and if a license conflict existed in a package. Protecode has a database containing millions of files from many open source projects hosted on several forges. When scanning the downloaded code, Protecode generates signatures and hashes that it compares against signatures of the files stored in Protecode's database. In this manner, Protecode's tools can identify if there are any matching files thereby indicating a file or part of the file exists in an open source project. Protecode also stores information regarding copyright and licenses of the open source projects found in the database, which will help identify any license conflicts between the open source components identified in the scanned code.

## Results

Across the 255 projects where licenses could be determined, 24% had permissive licenses. Four packages out of the 61 permissively licensed projects were confirmed of being contaminated with GPL code. GPL contamination was confirmed by checking if the permissively licensed package contained a file with a GPL attribution (i.e., a GPL license within the file or a reference to a GPL license within the file). The cases of GPL contamination include permissively licensed packages that included one or more GPL licensed files or including whole GPL licensed packages (*.js, *.py, *.tar).

We also found other cases where GPL contamination might have occurred, but it could not be confirmed with high certainty. For example, two permissively licensed projects may have inherited GPL code (modified or un-modified) and the GPL code does not contain a GPL reference within it. In another case, we found information on the project website that claimed

## License Compliance in Open Source Cybersecurity Projects

*Ahmed Shah, Selman Selman, and Ibrahim Abualhaol*

that a particular package was permissively licensed; however, when we downloaded the package, we found that the license was, in fact, restrictive.

Figure 1 compares the permissively licensed packages that are GPL contaminated with those that are not contaminated using a cluster plot of the total lines of unique code versus lines of third-party code. The figure shows that contaminated projects each have over 10,000 lines of third-party code and over 1,000 lines of unique code, although no other pattern is evident in this dataset, which contains only four cases of contamination.

*Package licenses*

Out of the 255 projects for which licensing information was available, 61 (24%) were found to have permissive licenses (i.e., MIT, BSD, or Apache). BSD-licensed projects were most common, accounting for nearly a third of permissively licensed projects and highlighting the flexibility inherited in this license. The MIT and Apache licenses were also common, each accounting for about 15% of the remaining permissively licensed projects. Of the permissively licensed packages contaminated with GPL-licensed code, two were licensed under BSD, one was licensed under MIT, and one contained a mix of permissive licenses.

The other 194 (76%) of 255 projects for which licensing information was available included a restrictively licensed (i.e., GPL) projects. One package was found to have a EUPL 1.1 license, which contained files that alluded to being GPL licensed. This package was grouped into the restrictive license category. Also included where packages were licensed under an LGPL (v3, v2, or v2.1), which could be considered moderately restrictive.

Figure 2 plots the number of programming languages used in each project against the number of lines of code for permissively and restrictively licensed packages. Figure 2 shows that packages with more than 1,000 lines of code are likely using one, two, or more languages, whereas packages with over 100,000 lines of code are likely using two or more programming languages. The GPL contamination boxes show the location of permissively licensed packages with GPL contamination. Out of the sample of 344 projects (which included projects that had licenses that could not be confirmed), the three most commonly used programming languages were C, Python, and PERL. This distribution of the four cases of GPL-code found in permissively licensed packages shows that contamination can occur regardless of the number of languages used.

**Figure 1.** Cluster plot of projects with permisive package licenses by lines of unique code and likely third-party code

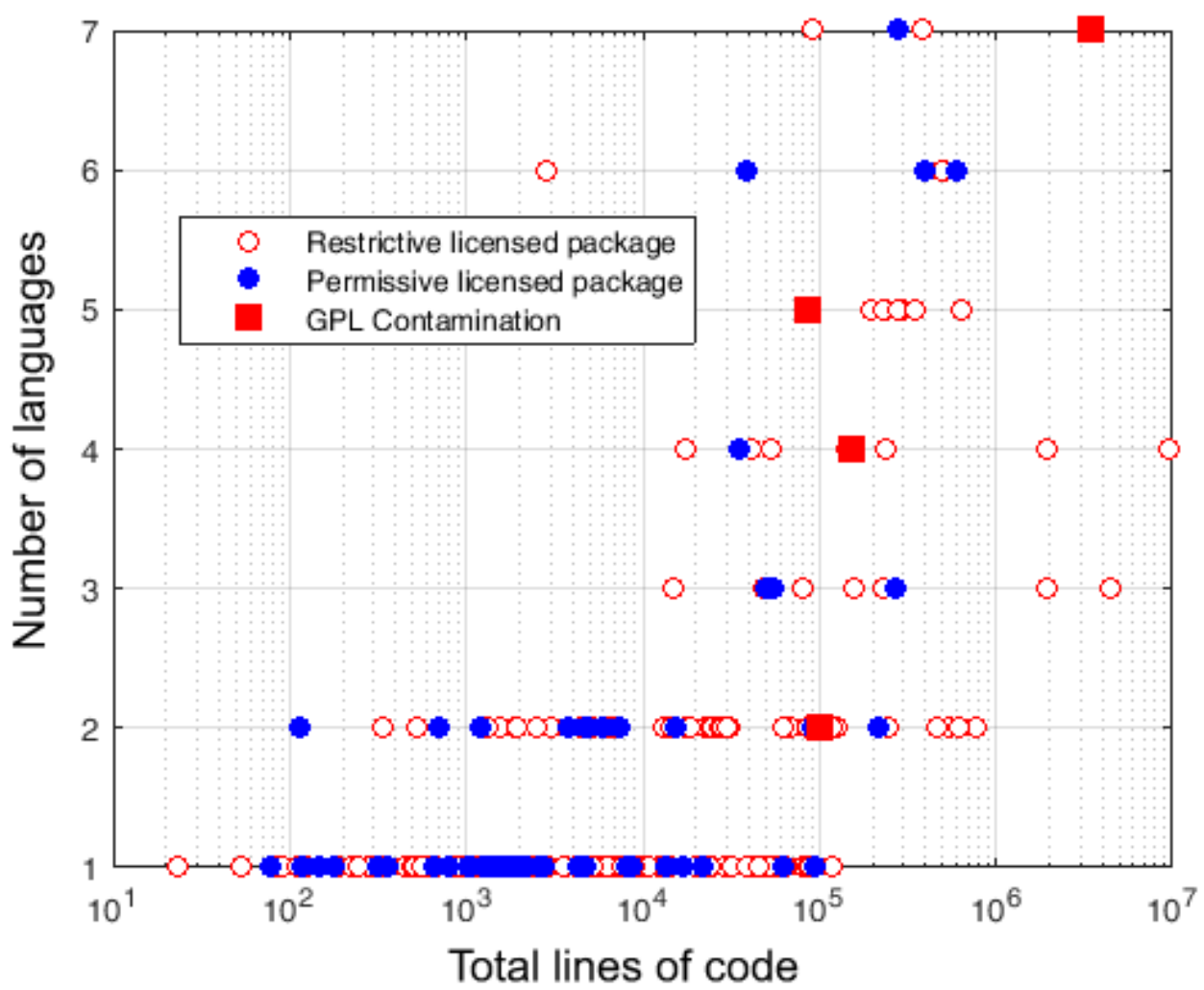

**Figure 2.** Programming languages versus total lines of code in restrictively licensed and permissively licensed packages, including evidence of GPL contamination in permissively licensed packages

# License Compliance in Open Source Cybersecurity Projects

*Ahmed Shah, Selman Selman, and Ibrahim Abualhaol*

*Copyright and license information in code files*

Often, when open source code is brought into projects, what is inherited is not the entire package of another project, but only a code snippet or a file from that project. If intellectual property claims (copyright) or licenses are not embedded within the file, there is a risk that the file could be mistakenly be used in a permissively licensed project, and this mistake could then be propagated into other projects, leading to viral contamination. In our dataset of 334 packages, in which we found 151,187 files (not including binaries), 39% of the files had no copyright information or did not refer to a license. For the files that did have either copyright or license information, 2% percent only made reference to a license, 43% made reference to a license and contained copyright information, and 16% only had copyright information (Table 1). Out of the 45% that did refer to a license, 63% of the files made reference to GPL, and 13% were standalone (not mixed) Apache, BSD, or MIT licensed.

*Volume of third-party code*

Protecode Enterprise Server 4 was used to determine the amount of third-party code that likely exists in each project in our subsample of 243 projects. When the Protecode software scans a file, it compares it against its database of known third-party code. If the Protecode software provided a suggested best match of third party for a file, for the sake of this article, we treat the entire file as third-party code.

Figure 3 presents the distribution of lines of code across projects, highlighting the third-party code and also the permissively licensed packages that are contaminated with GPL material. Figure 4 shows the distribution of projects by the percentage of the code within the project that is likely from a third party. Around 145 projects contain 0% to 10% third-party code while around 20 projects contain 90% or more third-party code. Across all projects, the average volume of third-party code is 27%.

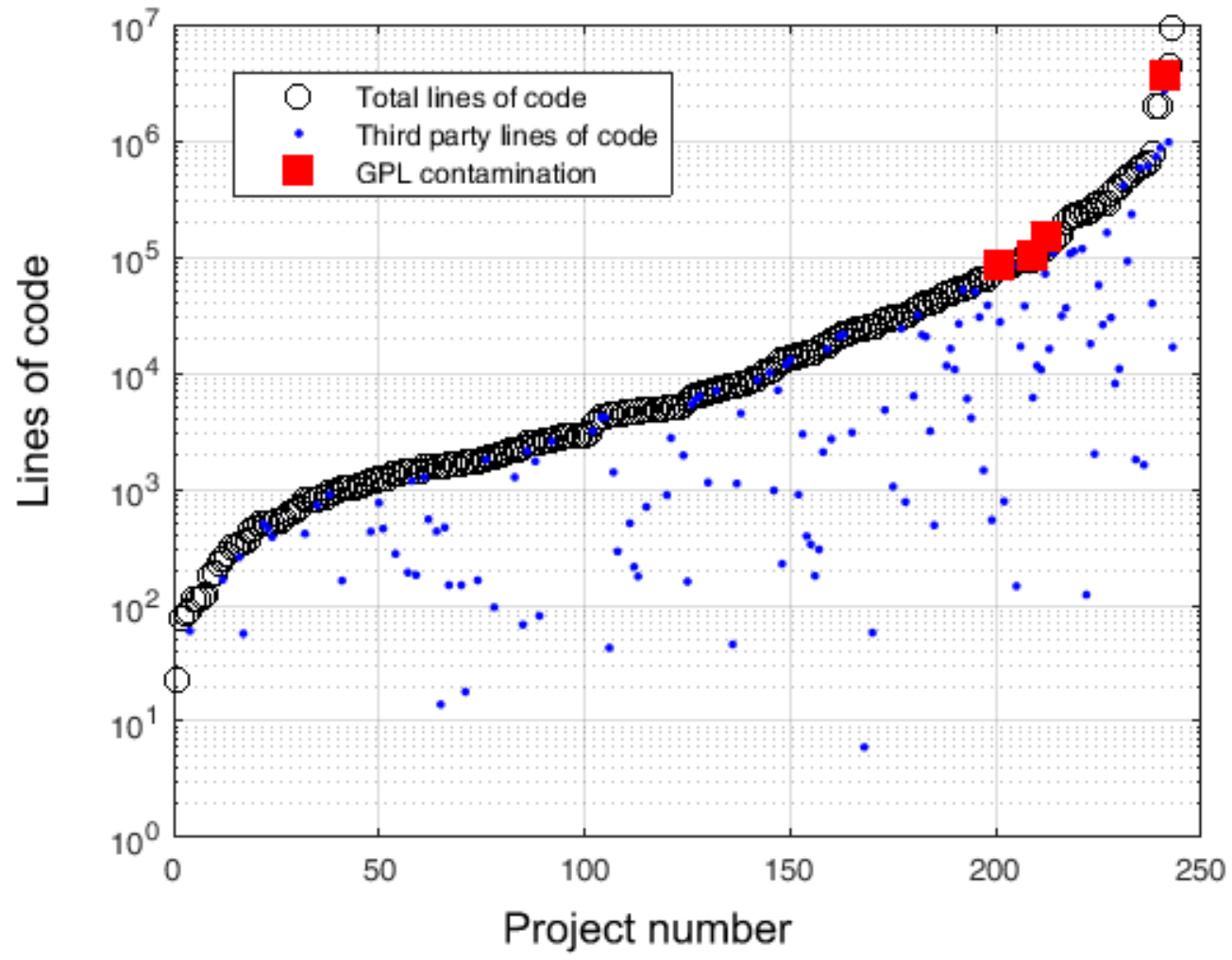

**Figure 3.** Extent of third-party code in the 243 sampled projects, ordered by total lines of code in each project

**Table 1.** Copyright and license information in 334 open source cybersecurity projects

<table>
<tr><td colspan="4">Full Dataset of 334 Projects:<br>151,187 files in total not including binaries</td></tr>
<tr><td colspan="3">IP or License Claim<br><br>Reference to license(s) or copyright<br><br>91,651 Files<br>61%</td><td rowspan="2">No IP or License Claim<br><br>No reference to copyright or license<br><br>59,536 Files<br>39%</td></tr>
<tr><td>Reference to license(s) only<br><br>3,267 Files<br>2%</td><td>Reference to license(s) and copyright<br><br>64,361 Files<br>43%</td><td>Reference to copyright only<br><br>24,023 Files<br>16%</td></tr>
</table>

# License Compliance in Open Source Cybersecurity Projects

*Ahmed Shah, Selman Selman, and Ibrahim Abualhaol*

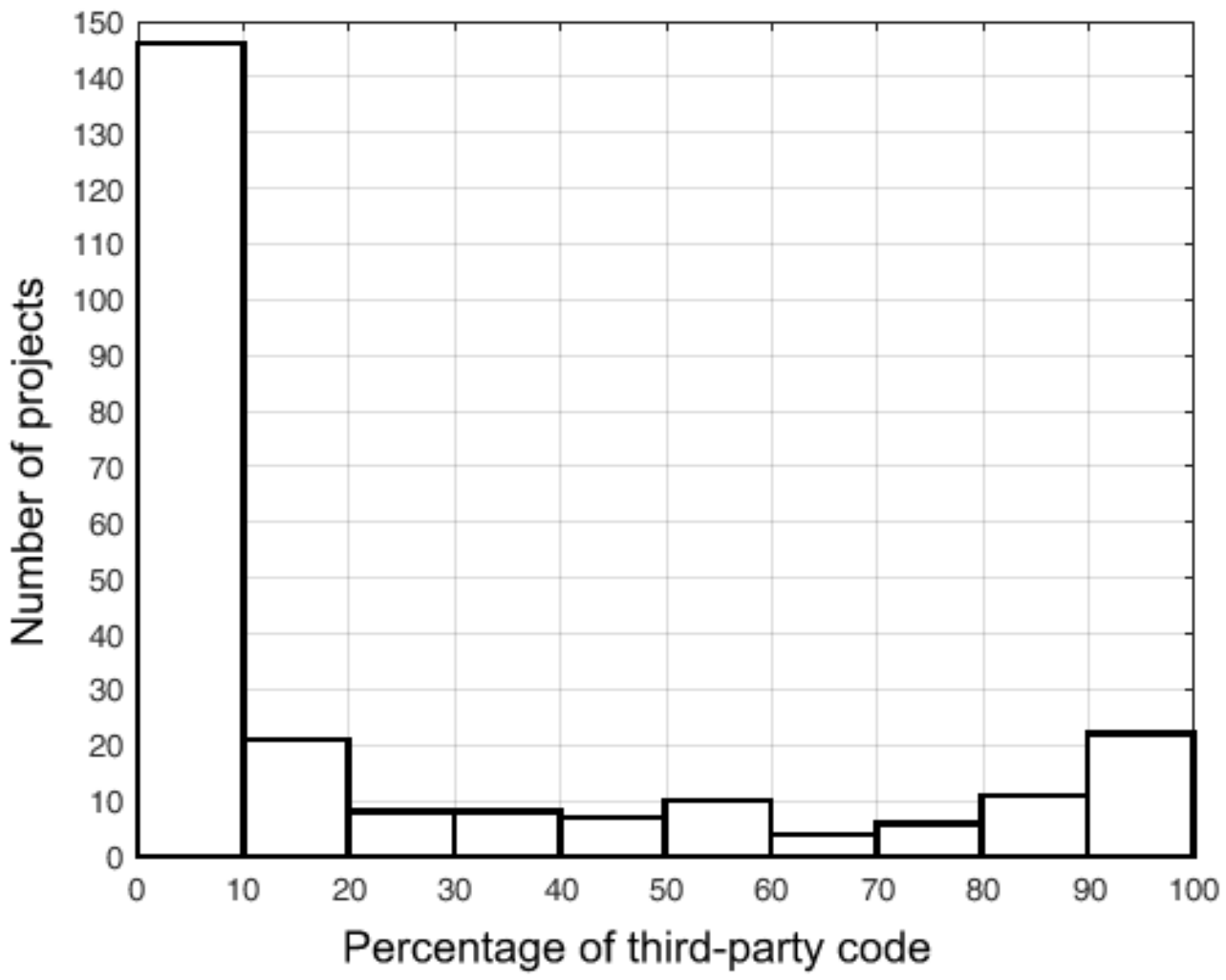

**Figure 4.** Histogram of number of projects versus the portion of package that is likely third-party code

## Future Work

This article provided an initial, exploratory analysis of open source cybersecurity projects to provide insight on open source license conflicts. Our results provide developers with insights into the characteristics of open source cybersecurity projects in terms of lines of code, languages used, and license types. In addition, we tried to identify the risk of permissively license projects being contaminated with GPL and the extent to which developers are adding copyright and license references to their code.

Future work could include statistical correlation analysis between different attributes, investigating a greater number of attributes (e.g., the number of contributors), and analyzing more projects to increase the power of the analysis in terms of detecting or ruling out the appearance of cluster patterns. Such work could lead to a classification of contamination probabilities based on a k-Nearest Neighbour (KNN) Algorithm.

## Conclusion

We found that the open source cybersecurity community is not adding copyright information or license references to files to claim intellectual property rights: 39% of files did not have copyright or license attribution. We suggest that managers should implement policies of adding copyright and licenses to their source code to ensure that intellectual property rights are claimed and to also make sure that GPL source code might not accidentally be consumed and contaminate a commercial product. We also found that there is no guarantee that packages with permissive licenses are not contaminated with restrictive licensed material: four out of 61 permissively licensed projects were contaminated with restrictive licenses. In addition, 76% of open source cybersecurity projects had restrictive package licenses and 24% had permissive package licenses. These findings suggest that the options for reusing open source code in the cybersecurity space are small with respect to selling proprietary software. However, the majority of restrictive licenses can be monetized through complementary services of open source products. Although much of the existing literature discusses the issue of open source licensing, licensing conflicts, and licensing compatibility, these studies are often light on data. In this study, we examined a dataset of over 300 open source cybersecurity projects and provides a steppingstone for further investigation in the open source cybersecurity domain. Although our findings revealed only four cases of contamination across 344 open source cybersecurity projects, the potential ramifications of such contamination for those individual warrant further study into how companies can mitigate this risk.

# License Compliance in Open Source Cybersecurity Projects

*Ahmed Shah, Selman Selman, and Ibrahim Abualhaol*

## About the Authors

**Ahmed Shah** holds a BEng in Software Engineering and is pursuing an MASc degree in Technology Innovation Management at Carleton University in Ottawa, Canada. Ahmed has experience working in cybersecurity research with the VENUS Cybersecurity Corporation and has experience managing legal deliverables at IBM.

**Selman Selman** is a Software Engineer at Synopsys under the Software Integrity Group. He is also carrying out graduate studies in Technology Innovation Management at Carleton University in Ottawa, Canada.

**Ibrahim Abualhaol** holds BSc and MSc degrees in Electrical Engineering from Jordan University of Science and Technology, an MEng in Technology Innovation Management from Carleton University in Ottawa, Canada, and a PhD in Electrical Engineering from the University of Mississippi in Oxford, United States. He worked for two years as a Wireless Engineer at Broadcom Corporation and as a System Engineer Intern at Qualcomm Incorporation in the United States. He then worked as an Assistant Professor of Wireless Communications at Khalifa University, United Arab Emirates for four years. Currently, he is a Cybersecurity R & D Engineer working on operationalizing collective intelligence with artificial intelligence to improve cybersecurity. He is senior member of IEEE, a member of Phi Kappa Phi, and a member of Sigma Xi.

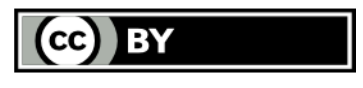